\documentclass[aps,twocolumn,showpacs,groupedaddress,nofootinbib]{revtex4-2}
\usepackage{epsfig}
\usepackage{bm}
\usepackage{amsmath,amssymb,times}
\usepackage{url}

\usepackage[colorlinks=true, pdfstartview=FitV, linkcolor=red, citecolor=blue, urlcolor=blue]{hyperref}

\newcommand{\UNIT}[1]{\mbox{$\,{\rm #1}$}}

\newcommand{\pvec}{\vec{p}\,}

\newcommand{\be}{\begin{equation}}
\newcommand{\ee}{\end{equation}}
\newcommand{\ba}{\begin{eqnarray}}
\newcommand{\ea}{\end{eqnarray}}

\newcommand{\fac}{\frac{\kappa}{(2\pi)^{3}}}

\newcommand{\nld}{{\cal D}}

\newcommand{\nldl}{\overleftarrow{{\cal D}}}
\newcommand{\nldr}{\overrightarrow{{\cal D}}}

\newcommand{\pspace}{\int\limits_{|\pvec|\leq p_{F_{b}}}\!\!\!\!\!\! d^{3}p}

\newcommand{\formbnm}{\frac{\Lambda_{{1}}^{2}}{ \Lambda_{{2}}^{2} + p^{2} } }

\newcommand{\overbar}[1]{\mkern 1.5mu\overline{\mkern-1.5mu#1\mkern-1.5mu}\mkern 1.5mu}

\begin{document}

\preprint{02-02}

\title{A novel solution to the hyperon-puzzle in neutron stars}
\author{Arsenia Chorozidou}
\author{Theodoros Gaitanos}
\email{tgaitano@auth.gr}
\affiliation{Physics Department, School of Physics, Aristotle University of Thessaloniki, 
54124 Thessaloniki, Greece}
\date{\today}

\begin{abstract}
Neutron stars offer a great opportunity to study highly compressed hadronic matter 
experimentally and theoretically. However, the so-called hyperon-puzzle arises at 
neutron star densities. The hyperon coexistence with other particles in compressed 
matter softens the equation of state and many widely-accepted models fail 
to reproduce precise observations of large neutron star masses. Here, 
we propose a novel mechanism to retain the stiffness of the high density state 
with hyperons by considering the 
explicit momentum dependence of their in-medium potentials. Our approach modifies 
conventional strangeness threshold conditions and generates new threshold effects 
on hyperons in high-density matter. We demonstrate these effects within the Non-Linear 
Derivative model, which incorporates baryon momentum-dependent fields based on 
empirical and microscopic studies. It turns out that even soft momentum-dependent 
strangeness fields do prohibit their populations in neutron star matter. The generic 
momentum dependence of strangeness potentials, as modeled by the non-linear 
derivative approach, is crucial for resolving the long-standing hyperon-puzzle 
in neutron stars.
\end{abstract}
\maketitle


\textit{Introduction --} 
The discovery of massive neutron stars~\cite{Demorest:2010bx,Antoniadis:2013pzd,NS0,NS1,NS2} 
has triggered debates in the nuclear/hadron physics and astrophysics communities 
about the properties of hadrons at high baryon densities and 
isospin asymmetries~\cite{Steiner:2012xt,Lattimer:2021emm}, 
particularly related to the 
hyperon-puzzle~\cite{Lonardoni:2014bwa,Gerstung:2020ktv,Vidana:2010ip,Weissenborn:2011ut}. 
At high energy densities, baryons with strangeness content, known as hyperons, 
can be produced. They soften the hadronic equation of state (EoS) of compressed 
neutron star (NS) matter and reduce the maximum NS mass below astrophysical 
observations. Thus, various phenomenological and microscopic nuclear matter 
approaches have failed to reproduce the observed large neutron star (NS) masses when 
including strangeness in their descriptions. Proposed solutions include modified 
vector repulsion of baryons at high densities and the inclusion of additional 
strangeness exchange mesons in phenomenological 
approaches~\cite{Bednarek:2012, Weissenborn:2012, Oertel:2015, Maslov:2015},
while microscopic 
approaches~\cite{Gerstung:2020ktv,Petschauer:2016,Haidenbauer:2023qhf,Fujiwara:2006yh,Logoteta:2019utx,Chatterjee:2015pua} 
predict an explicit momentum dependence (MD) of hyperon potentials 
based on scattering data. 

Inspired by the microscopic MD of in-medium hyperon potentials and the successful 
predictions of the Non-Linear Derivative (NLD) model 
for nuclear 
matter systems~\cite{Gaitanos:2013,Gaitanos:2011yb,Gaitanos:2015xpa}, 
we extend the NLD approach to $\beta$-equilibrated matter with strangeness degrees of freedom. 
The explicit MD of the hyperon potentials in the NLD model manifests new effects 
on their threshold conditions. Hyperons can disappear even when the threshold 
condition is met at zero momentum. Moreover, a soft MD of strangeness 
potentials may not allow their population in NS matter. We discuss these 
new momentum-dependent threshold effects within the NLD model, which is based on hyperon 
potentials in the spirit of the chiral effective field theory ($\chi$-EFT). 
We conclude that within the NLD approach the hyperon-puzzle issue is resolved when 
considering explicitly momentum-dependent potentials.

\textit{The NLD model --}
The NLD formalism~\cite{Gaitanos:2013} is based on the 
conventional Relativistic Hadro-Dynamics (RHD)~\cite{Serot:1984ey, Boguta:1977xi}
with the usual free Lagrangians for the baryons 
$\Psi_{b}$ ($b=N, Y$ with $Y$ denoting the hyperons) 
and those for the exchange isoscalar-scalar, isoscalar-vector and isovector-vector meson 
fields $m=\sigma,\omega,\rho$. 
The NLD interaction Lagrangian exhibits the 
same structure as that in the RHD model, but with non-linear derivative contributions 
incorporated between the bilinear baryon fields as 
\begin{equation}
{\cal L}_{int}^{m} = \sum_{b} \frac{g_{m b}}{2}
	\left[
	\overbar{\Psi}_{b}
	 \nldl_{b} \Gamma_{m}
	\Psi_{b}\,\varphi_{m}
	+ \varphi_{m}\overbar{\Psi}_{b}\Gamma_{m}
	 \nldr_{b}
	\Psi_{b}
	\right]
	\,,
\end{equation}
with obvious couplings $g_{mb}$ and Lorentz-factors 
$\Gamma_{m}=\UNIT{1},\gamma^{\mu},\vec{\tau}\gamma^{\mu}$ for the various vertices 
involving the exchange meson fields $\varphi_{m}$. 
The baryonic non-linear derivative operators $\nld_{b}$ act on the baryon fields and  
extend the conventional coupling scheme between baryons  
and mesons. In momentum space, they are momentum-dependent functions 
of a common monopole-like form with appropriate cut-off parameters~\cite{Gaitanos:2013}. 
They regulate the MD of all baryon potentials in consistency with all 
available empirical and microscopic knowledge. In infinite 
matter and within the relativistic mean-field approximation the NLD approach yields 
quasi-free Dirac equations for baryons, featuring explicitly momentum-dependent 
scalar and vector selfenergies. For instance, the vector component 
is given by 
\begin{align}
\Sigma^{\mu}_{b}(p)  =  g_{\omega b} \, \omega^{\mu} \, {\cal D}_{b}(p)
+\tau_{3b} g_{\rho b}  \,  \rho^{\mu} \, {\cal D}_{b}(p)
\label{Sigmav_B}
\,,
\end{align}
with the isospin factor $\tau_{3b}$ and the 
regulator ${\cal D}_{b}(p)$ for a baryon $b$. It depends implicitly on the total 
baryon density $\rho_{B}$ too. Similar expressions occur 
for the scalar selfenergies. For simplicity, the scalar and time-like component  
of the baryon selfenergy will be denoted as $S_{b}$ and $V_{b}$, respectively, 
omitting their explicit MD in most of the expressions. 
The NLD equations of motion for the $\sigma$ meson and for the time-like component 
of the $\omega$ field read as
\begin{align}
m_{\sigma}^{2}\sigma + \frac{\partial U}{\partial\sigma} = & 
\sum_{b} 
g_{\sigma b} \fac \!\!\!\!\!\!
\pspace 
\frac{m^{*}_{b}}{E^{*}_{b}} \, \nld_{b}(p)
\label{sigma}
\\
m_{\omega}^{2}\omega = & 
\sum_{b} 
g_{\omega b}\fac \!\!\!\!\!\!
\pspace \, \nld_{b}(p)
\label{omega}
\,,
\end{align}
with the meson masses $m_{\sigma,\omega}$, the usual self-interaction 
term $U=U(\sigma)$~\cite{Boguta:1983}, the 
spin degeneracy factor $\kappa=2$, the effective mass $m^{*}_{b}(p)=M - S_{b}(p)$ 
and the in-medium energy $E^{*}_{b}(p)=\sqrt{m^{*2}(p)+p^{2}}$. 
An expression similar to Eq.~(\ref{omega}) occurs for the isovector meson $\rho$. 
Finally, the EoS is obtained from the energy density 
\begin{align}
\varepsilon = T^{00} = 
\sum_{b} \fac \pspace \, E_{b}(p) - \langle{\cal L}\rangle
\label{tensor}
\end{align}
with the baryon energy determined from the quasi-free dispersion relation 
\begin{align}
E_{b}(p) = \sqrt{m^{*2}(p)+p^{2}} + V_{b}(p)
\label{BarEnergy}
\end{align} 
self-consistently due to the explicit momentum dependencies. 

Note that the regulators ${\cal D}_{b}$ enter not only in the baryon selfenergies 
explicitly, Eq.~(\ref{Sigmav_B}), but they show up in the meson-field equations 
implicitly too, Eqs.~(\ref{sigma},\ref{omega}). 
Thus, within the NLD model, the scalar 
and, in particular, the vector fields are suppressed (or regulated) with 
increasing density in a non-linear manner. This in-medium vector suppression was 
necessary for an adequate description of the EoS and, at the same time, of the 
in-medium optical potentials~\cite{Gaitanos:2013}. 

We focus now on the NLD effects to the NS matter including hyperons, that is, 
hadronic matter in $\beta$-equilibrium. 
Although there are relatively precise empirical analyses available for 
the MD of the in-medium proton optical potential, 
the situation in the strangeness sector is still lacking. 
In particular, available data in the single-strangeness sector 
allow a theoretical consensus concerning the MD of the single-strangeness 
($\Lambda,\Sigma^{0,\pm}$)-potentials~\cite{cons1} 
at matter densities close to saturation. However, 
in the double-strangeness sector the theoretical uncertainties for the 
the $\Xi^{-,0}$-potentials at finite momentum are still too large~\cite{cons1,cons2}. 
They can lead to a quite ambiguous MD at high densities. 
A more systematic study concerning the double-strangeness sector is 
necessary and goes beyond the scope of the present work. 
Therefore, in order to keep the presentation of the new strangeness threshold 
effects as transparent as possible, we consider the ($\Lambda,\Sigma^{0,\pm}$) 
hyperons in this work. 
We use recent microscopic $\chi$-EFT potentials as a reference~\cite{Petschauer:2016}. 
The initial NLD results for hyperons were presented in~\cite{Gaitanos:2021}. 
We have improved and expanded upon them to encompass conditions that 
are significant for NS matter. 
The NLD hyperon parameters as shown in Table~\ref{tab1} are the following: the 
Lorentz-scalar, isoscalar factor $\chi_{\sigma}$ defined by $g_{\sigma Y}=\chi_{\sigma}g_{\sigma N}$ 
($\chi_{\omega,\rho}$ are fixed by SU(6)). The cutoffs for the various hyperons, which enter 
into the regulators $D_{Y}(p)=\formbnm$ for the various $\sigma$-, $\omega$- and 
$\rho$-hyperon interactions~\cite{Gaitanos:2021}. 
\begin{table*}
\begin{tabular}{|cccc|ccccc|cccc|ccccc|}
\hline
\multicolumn{4}{|c|}{$\Lambda$ hyperon} & \multicolumn{5}{|c|}{$\Sigma^{-}$ hyperon} & 
\multicolumn{4}{|c|}{$\Sigma^{0}$ hyperon} & \multicolumn{5}{|c|}{$\Sigma^{+}$ hyperon} \\
$\chi_{\sigma}$ & $\Lambda_{\sigma}$ & $\Lambda_{\omega_{1}}$ & $\Lambda_{\omega_{2}}$ & 
$\chi_{\sigma}$ & $\Lambda_{\sigma}$ & $\Lambda_{\omega_{1}}$ & $\Lambda_{\omega_{2}}$ & $\Lambda_{\rho}$ & 
$\chi_{\sigma}$ & $\Lambda_{\sigma}$ & $\Lambda_{\omega_{1}}$ & $\Lambda_{\omega_{2}}$ & 
$\chi_{\sigma}$ & $\Lambda_{\sigma}$ & $\Lambda_{\omega_{1}}$ & $\Lambda_{\omega_{2}}$ & $\Lambda_{\rho}$   \\
\hline
 0.83 & 0.76 & 0.95 & 0.79 & 0.63 & 0.85 & 0.95 & 0.79 & 0.6 & 0.6 & 0.8 & 0.95 & 0.79 & 0.65 & 0.62 & 0.95 & 0.8 & 0.6 \\ 
 0.91 & 0.76 & 0.95 & 0.75 &  &  &  &  &  &  &  &  &  &  &  &  &  &  \\ 
\hline
\end{tabular} 
\caption{Hyperon parameters: the $\sigma$-hyperon scaling factors $\chi_{\sigma}$ 
and the cutoff parameters (in units of GeV) 
$\Lambda_{\sigma}$ ($\Lambda_{1}=\Lambda_{2}=\Lambda_{\sigma}$), 
$\Lambda_{\omega_{1,2}}$ and $\Lambda_{\rho}$ ($\Lambda_{1}=\Lambda_{2}=\Lambda_{\rho}$) 
for the various hyperons. 
The two parameter sets for the $\Lambda$ hyperon define the band limits in 
Fig.~\ref{fig1}(left panels). }
\label{tab1}
\end{table*}


\textit{Momentum-dependent strangeness thresholds --}
The key observable for understanding better the strangeness threshold conditions 
in $\beta$-equilibrium and for comparing different models is the 
Schr\"{o}dinger-equivalent optical (or simply optical) potential given by
\begin{align}
U_{opt}^{b} = -S_{b} + \frac{E_{b}}{m_{b}}V_{b} + 
\frac{1}{2m_{b}}\left( S_{b}^{2} - V_{b}^{2}\right)
\label{Uopt}
\,.
\end{align}
Its real part describes the hadronic mean field felt by the baryon $b$ with a momentum 
$p=|\vec{p}\,|$ relative to the hadronic matter at rest at a given baryon density 
$\rho_{B}$. 
For the theoretical description of NS matter we consider the NS composition of the baryons 
$b=N, \Lambda, \Sigma^{0,\pm}$ and of the electrons as the only leptonic contribution. 
The electrons are treated as an ideal Fermi-Gas. 
Imposing charge neutrality, vanishing strangeness chemical potential 
due to the infinite time scale of a NS relative to the weak interaction 
time scale and $\beta$-equilibrium,  
the baryon chemical potentials, $\mu_{b}$, and those of the electrons, $\mu_{e}$, 
are related through the chemical equilibrium conditions~\cite{Glendenning:1985}
($q_{b}$ is the charge of a baryon $b$)
\begin{align}
\mu_{b} = \mu_{n} - q_{b}\,\mu_{e},~~\mu_{b} = \sqrt{p_{F_{b}}^{2}+m^{*^{2}}_{b}}+V_{b}
\label{betaEq}
\,,
\end{align}
Together with the total baryon density conservation and charge neutrality constraints, 
\begin{align}
\rho_{B} = \sum_{b}\rho_{b},~~\sum_{b}q_{b}\rho_{b} - \rho_{e} = 0
\label{conserv}
\,,
\end{align}
one has to solve self-consistently a set of the 
non-linear equations~(\ref{sigma},~\ref{betaEq},~\ref{conserv}) for the $\sigma$-meson and 
the two independent chemical potentials $\mu_{n}$, $\mu_{e}$. At each iteration 
step the Fermi-momenta of each particle are calculated from the chemical equilibrium conditions 
in Eq.~(\ref{betaEq}) as the positive real solutions, 
that is, the equation $\mu_{b}=E_{b}(p)$ with the solution being the 
Fermi-momentum $p_{F_{b}}$ for a baryon $b$. 
Eqs.~(\ref{betaEq}) 
indicate the threshold 
conditions for the 
particles heavier than the neutron, 
that is, the strangeness thresholds. 
Therefore we focus the following discussion on the hyperons (Y). 
In the NLD model, however, the explicit MD of the effective mass $m^{*}_{b}$ 
and of the vector selfenergy $V_{b}$ 
will induce modifications 
of the conventional threshold conditions with new upcoming effects for hyperons. 

At first, without any explicit MD in the potential, 
when the chemical potential $\mu_{n}-q_{Y}\mu_{e}$ exceeds hyperon's lowest 
energy, that is
\begin{align}
\mu_{Y}=\mu_{n}-q_{Y}\mu_{e} > E_{Y}(0)
\,,
\label{thresh-std}
\end{align}
a hyperon $Y$ will be produced with a finite Fermi-momentum $p_{F_{Y}}$ as the solution 
of the equation $\mu_{Y}=E_{Y}(p_{F_{Y}})$. The inequality~(\ref{thresh-std}) is the conventional 
threshold condition~\cite{Glendenning:1985}.

With an explicit MD, however, 
the hyperon selfenergies (or the optical potential) entering 
in the hyperon energy $E_{Y}(p)$ depend explicitly on the hyperon momentum. This generic 
MD of the fields can induce a non-trivial momentum dependence of the hyperon in-medium 
energy $E_{Y}(p)$ which may differ from the usual monotonically increasing $p$-behavior. 
In other words, 
the soft/stiff nature of the optical potential versus the momentum at a 
fixed $\rho_{B}$ influences the stiffness of the  
hyperon in-medium energy 
as function of momentum. 
Consequently, it affects the Fermi-momentum 
value as the solution of $\mu_{Y}=E_{Y}(p_{F_{Y}})$, 
if such a solution exists. 
Thus, a stiff-like momentum-dependent behavior of the 
hyperon energy 
will likely shift the threshold to higher densities, as expected. 
A weakly soft-like MD of the fields will 
likely increase the strangeness population, again as expected. However, 
a particular attention shall be given to the case of a very soft 
momentum dependent fields. 
They do reveal new effects on the strangeness thresholds. 
Very soft momentum-dependent fields may cause 
a finite hyperon population even if the threshold condition, 
Eq.~(\ref{thresh-std}), is not fulfilled. 
This is an extreme case 
and does not occur in the calculations. On the other hand,  
a very soft MD of the selfenergies can eventually prohibit 
the hyperon population even with a fulfilled threshold condition at vanishing momentum. 
This latter case occurs in the calculations and will be particularly discussed 
in the presentation of the NLD results below. 

\begin{figure}[t]
\begin{center}
\includegraphics[clip=true,width=1\columnwidth,angle=0.]
{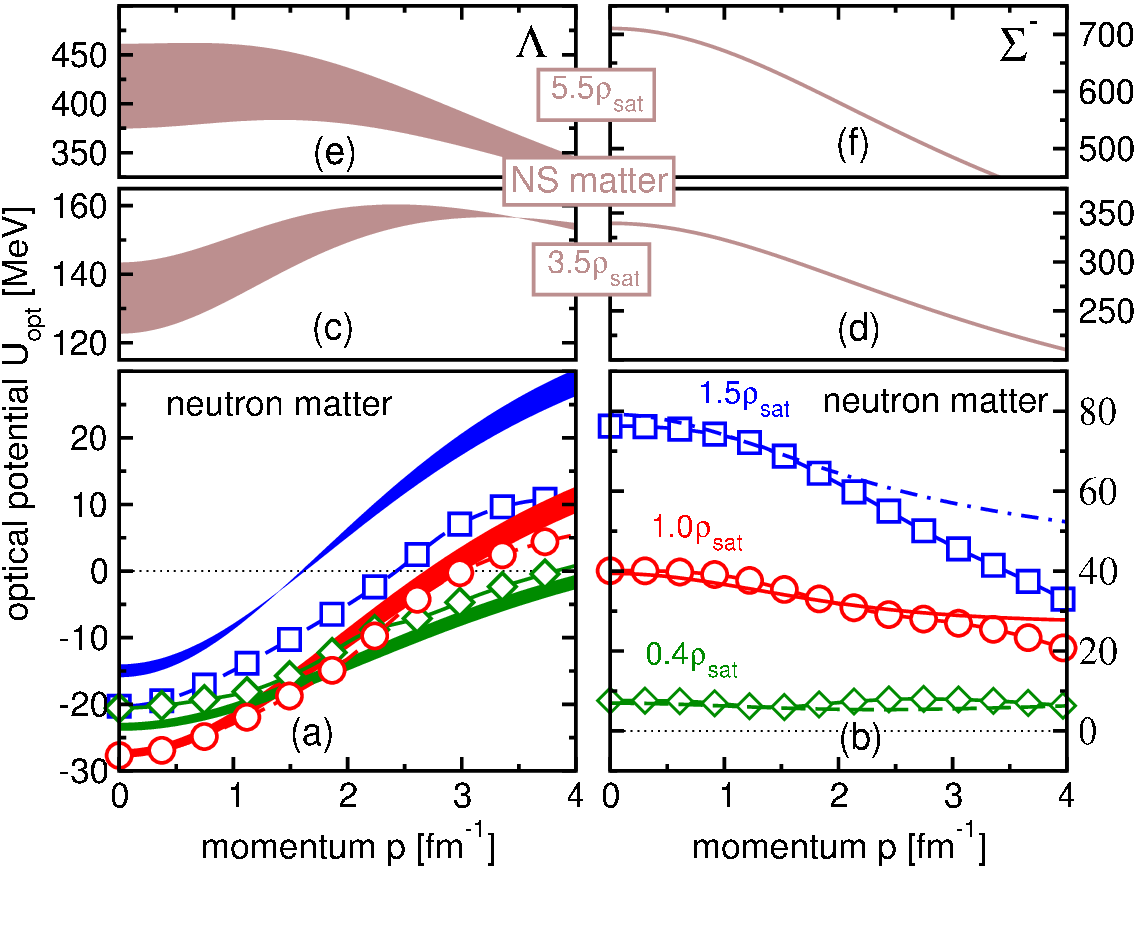}
\caption{\label{fig1} 
(Color Online) 
In-medium optical potentials $U_{opt}$ for $\Lambda$ (left) and $\Sigma^{-}$ (right) 
hyperons versus their momentum $p$. 
(a,b) Comparison between NLD (green, red and blue bands) and 
$\chi$-EFT optical potentials~\protect\cite{Petschauer:2016} 
(diamond-, circle- and square-symbols) for pure neutron 
matter at densities of $\rho_{B}=0.062, 0.155$ and $0.2325$ (in units of 
$fm^{-3}$), respectively. NLD predictions for  
NS matter ($\beta$-equilibrium) at high densities of 
$\rho_{B}=0.5~fm^{-3}$ (c,d) and $\rho_{B}=0.8~fm^{-3}$ (e,f) are shown too.
\vspace{-0.3cm}
}
\end{center}
\end{figure}

Therefore, the interplay between momentum dependencies and strangeness population 
in $\beta$-equilibrated compressed matter turns out to be more complex than one would 
expect. Not only a stiff MD can reduce or forbid the hyperon population, 
but a soft momentum behavior of the hyperon potential can, in general, 
prohibit the strangeness production in NS matter too.

\textit{NLD results for neutron star matter --}
An appropriate discussion about momentum-dependent threshold effects at high densities 
requires an adequate MD of the hyperon potentials at saturation density, which is 
better accessible in theory and eventually in future experiments. 
This has been realized within the 
NLD model in Ref.~\cite{Gaitanos:2021}. Here we have refined the MD of the hyperon 
potentials according to the corresponding $\chi$-EFT ones, 
see Table~\ref{tab1}. 
The comparisons between the NLD and the microscopic potentials are summarized 
in Fig.~\ref{fig1} (panels (a,b)). The NLD cutoffs were adjusted such to reproduce 
as close as possible the momentum dependence of the $\chi$-EFT 
potentials 
at the saturation density 
$\rho_{B}=0.155~fm^{-3}$ only. Although all the $\Lambda,\Sigma^{0,\pm}$-hyperons are 
included in the calculations, we restrict the discussion to the $\Lambda$ and $\Sigma^{-}$ 
baryons as the most prominent candidates for softening the NS EoS at 
high baryon densities. 

\begin{figure}[t]
\begin{center}
\includegraphics[clip=true,width=1.0\columnwidth,angle=0.]
{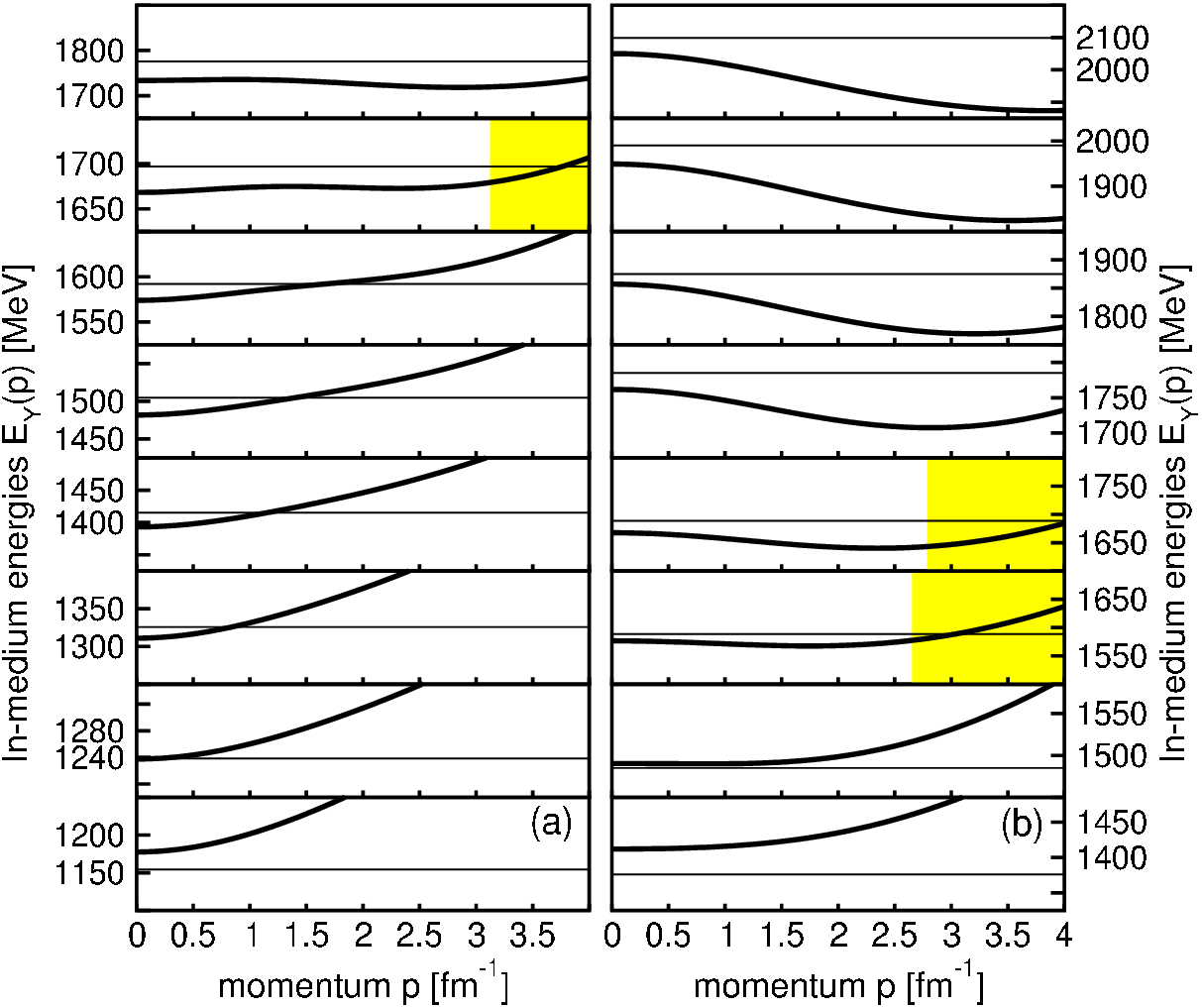}
\caption{\label{fig2} 
(Color Online) 
Demonstration of novel threshold effects for hyperons in the NLD model: 
the thick curves show the in-medium energies $E_{b}(p)$ 
for $\Lambda$ (a) and $\Sigma^{-}$ (b) hyperons versus their momentum $p$ 
for NS matter at baryon densities from $\rho_{B}=0.4~fm^{-3}$ (bottom) up 
to $\rho_{B}=1.1~fm^{-3}$ (top) in steps of $0.1~fm^{-3}$. 
The thin lines correspond to the thresholds $\mu_{n}-q_{Y}\mu_{e}$. 
The shaded areas indicate the forbidden regions of momenta higher than the 
Fermi-value at a given baryon density. Even when 
the population threshold is exceeded, 
the hyperons don't show up in some cases. 
\vspace{-0.3cm}
}
\end{center}
\end{figure}

\begin{figure}[t]
\begin{center}
\includegraphics[clip=true,width=0.9\columnwidth,angle=0.]
{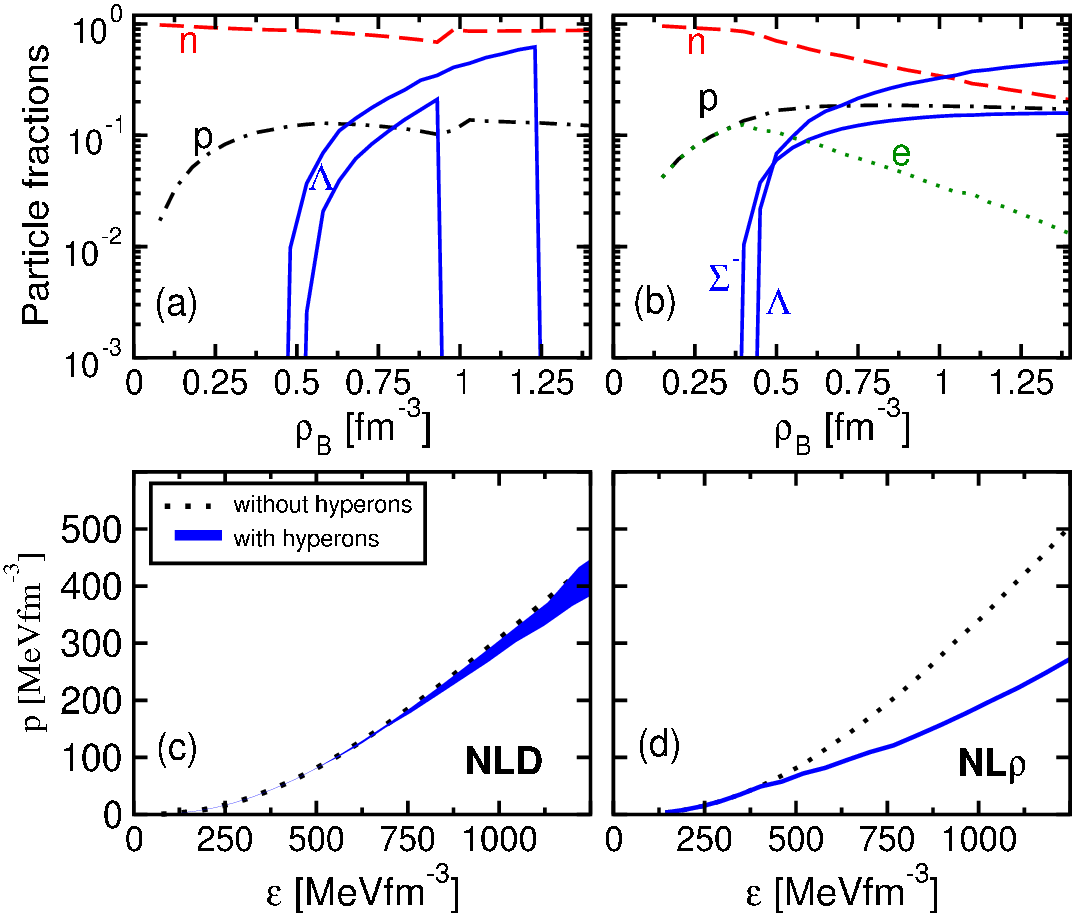}
\caption{\label{fig3} 
(Color Online) 
(a,b) Particle fractions, as indicated, versus the baryon density $\rho_{B}$. 
(c,d) the NS EoS in terms of the pressure ($p$) versus the energy ($\varepsilon$) 
densities. The results refer to the conventional NL$\rho$ model~\protect\cite{NLr} 
(b,d) and to the NLD approach (a,c) for NS matter 
without hyperons (dotted curves in (c,d)) and with the hyperons included (solid 
curve for NL$\rho$ and filled band for NLD). In the NLD model the electron-curve in (a) equals 
the p-curve and the $\Sigma$-hyperons are not produced. The lower and upper $\Lambda$-curves 
in (a) refer to the lower and upper NS EoS limits in (c).
\vspace{-0.5cm}
}
\end{center}
\end{figure}
At first, the NLD model gives a quite non-trivial behavior in density and 
momentum for the $\Lambda$- and the $\Sigma^{-}$-potentials in pure neutron matter, 
which is fully consistent with the $\chi$-EFT calculations at the low 
density region. More specifically, the NLD approach predicts an overall repulsive  
character of the $\Sigma^{-}$-potential with a soft MD. This is in perfect 
line with the microscopic calculations. This soft-like potential behavior 
versus the $\Sigma^{-}$-momentum is retained and becomes stronger 
at higher densities (panels (d,f) in Fig.~\ref{fig1}). 

The $\Lambda$-potential, on the other hand, does manifest a more complex MD with 
increasing densities up to the relevant NS region. 
We thus give more attention to the adjustment of the NLD 
$\Lambda$-potential to the microscopic results by using different cutoff 
choices. This results to the filled bands in Fig.~\ref{fig1} (panel (a)). 
Having in mind that all fits refer to the $\chi$-EFT results at saturation density only, 
the comparison between NLD and $\chi$-EFT $\Lambda$-potentials is fairly well at the 
given densities around saturation. It turns out that the NLD model predicts 
the microscopic in-medium behaviors of the hyperon potentials very well. 

The high density sector of the in-medium $\Lambda$-potential deserves a 
particular discussion, since it is the relevant region for NS matter. 
At first, note that the potential value at zero momentum for 
$\rho_{B}\approx 5\rho_{sat}$ 
is consistent with recent microscopic studies~\cite{Gerstung:2020ktv}. 
As a novel feature, a density-dependent transition from a 
stiff (Fig.~\ref{fig1}, panel (a)) to a soft (Fig.~\ref{fig1}, panels (c,e)) 
momentum-dependent behavior shows up in the $\Lambda$-potential. In particular, 
with increasing baryon density, the $\Lambda$-potential exhibits the expected 
repulsive character for all momenta, 
but with a decreasing tendency (a very soft behavior) as function of 
momentum. This is clearly 
visible at $\rho_{B}=5.5\rho_{sat}$ (panel (e) in Fig.~\ref{fig1}). This 
density-dependent stiffness transition in momentum originates from the NLD regulators. 
They show up explicitly in the selfenergies and implicitly in the vector-meson 
source term. Therefore, they soften significantly the MD of the strangeness potential 
with rising density. 

The non-trivial interplay of the hyperon potentials in density and, in 
particular, in momentum is manifested in the 
in-medium energies 
in Fig.~\ref{fig2}. 
Note again the explicit MD of the selfenergies entering into the hyperon 
in-medium energy. 
The $\Lambda$ in-medium energies 
with the stiff MD at the low NS density region 
of $\rho_{B}=0.4$ fm${}^{-3}$ 
cross the corresponding thresholds at momenta below the Fermi-momentum of the 
given baryon density. That is, they can be populated. 
The repulsive character of the $\Sigma^{-}$ chemical potential does not allow 
their population in this low NS density region.  

With increasing density, however, 
the momentum-dependent stiffness transition of the hyperon potentials sets in and 
changes the strangeness population drastically. 
Indeed, due to the very soft MD of the optical potential 
the $\Sigma^{-}$ in-medium energy 
does not exceed 
the corresponding threshold at momenta below the corresponding Fermi-momentum at these 
high densities, even if the threshold is fulfilled at vanishing momentum. 
No solution exists and they cannot be populated. 
The situation is similar for the $\Lambda$ hyperons. Here the more 
attractive nature together with the stiff MD of the $\Lambda$ 
in-medium energy 
allows still the $\Lambda$ population. However, with rising 
density the stiffness transition is more pronounced for the $\Lambda$ potential 
and prevents 
the production of the $\Lambda$ baryons again. 
Note that in some cases the in-medium energies 
do surpass the threshold line, however, at momenta higher than the 
allowed maximum value of the 
Fermi-momentum (indicated with the yellow areas in Fig.~\ref{fig2}). Thus, in these 
cases the $\Lambda$-hyperons cannot be populated. It turns out that,  
within the NLD model, the $\Sigma$-hyperons cannot be produced at all 
in NS matter. Only the $\Lambda$-hyperons can be populated, however, 
in a narrow density region.  

We discuss now the particle fractions and the NS EoSs within the NLD approach 
in comparison with a conventional relativistic mean-field (RMF) model. In general, 
any conventional RMF model results to an EoS softening for NS with hyperons. 
A stiffness restoration is possible by introducing additional vector-like 
selfinteraction terms or additional strange vector mesons with more additional 
parameters~\cite{Bednarek:2012, Weissenborn:2012, Oertel:2015, Maslov:2015,Providencia:2013}.  
For a meaningful comparison with the NLD approach we utilize the conventional 
NL$\rho$ model from Ref.~\cite{NLr}. It gives similar conditions, i.e., a 
similar NS EoS for nucleons and comparable hyperon potentials at low momenta 
relative to NLD.  In this way one can reveal better the novel  
features of the NLD approach. This is shown in Fig.~\ref{fig3} in terms of the 
particle fractions and NS EoSs. At first, the 
NL$\rho$ EoS is similar to the NLD EoS for NS matter without hyperons. 
On the other hand, it is clearly seen that the $\Lambda$- as well as the $\Sigma^{-}$-hyperons 
contribute significantly to the NS composition within the NL$\rho$ model. 
The NL$\rho$ hyperonic NS EoS is softened largely even with comparable hyperon potentials 
relative to NLD at low momenta. 
The situation within the NLD model is very different (panels (a,c) in Fig.~\ref{fig3}). 
The $\Sigma$-hyperons, in particular the energetically 
favored $\Sigma^{-}$-hyperons, are prohibited and cannot be populated. 
The $\Lambda$-hyperons are produced above a higher density threshold 
relative to the NL$\rho$ case, however, 
with much lower fraction. Note that they disappear again at higher densities due to 
the stiffness transition effect. 
As an important consequence, 
the NLD EoS does not get softened significantly in NS matter by including the hyperons. 
The maximum NS mass $M$ is $M\approx 2.05 M_{\odot}$ without hyperons. This value 
of $M\approx 2.05 M_{\odot}$ is maintained with the inclusion of hyperons in the NLD 
model.  

\textit{Summary and Conclusions --}
In summary, we proposed a solution to the persistent hyperon-puzzle in neutron stars. It 
is based on the in-medium strangeness MD, realized through the NLD model. 
This model successfully describes the non-trivial features of empirical and 
microscopic baryon in-medium optical potentials, and its application to neutron 
star matter with hyperons is appropriate. The NLD momentum-dependent hyperon fields 
generate novel effects on their threshold conditions
by preventing their population even when the requisite threshold 
conditions are met. 
By relying to the MD of the microscopic $\chi$-EFT 
calculations, the NLD model predicts NS matters  
with low $\Lambda$-hyperon fractions inside a limited density region only with no other 
hyperons present, resulting in a stiff strangeness NS EoS. 
It turns out that within the NLD approach the hyperon-puzzle issue is 
successfully resolved, particularly when considering momentum-dependent 
hyperon fields.

\section*{Acknowledgments}
We acknowledge Dr. K. Kosmidis for suggestions 
concerning the numerical treatments and Prof. J. Haidenbauer for valuable 
discussions and support with the microscopic calculations. 


\end{document}